\documentstyle[12pt]{article}
\font\mtbb=msbm10 scaled \magstep1

\def\l#1{\label{#1}}
\def\be{\begin{equation}}
\def\ee{\end{equation}}
\def\bea{\begin{eqnarray}}
\def\eea{\end{eqnarray}}
\def\ot{\otimes}
\def\={\; = \;}

\def\s{\sigma}

\def\bt{\bullet}

\def\a{{\cal A}}

\def\i{{\cal J}}
\def\m{{\cal M}}
\def\s{{\cal S}}
\def\rb{\hbox{\mtbb R}}
\def\cb{\hbox{\mtbb C}}

\newtheorem{lemma}{Lemma}[section]
\newtheorem{observation}{Observation}[section]

\begin{document}
\title{Noncommutative Differential Geometry with Higher Order Derivatives}
\author{Andrzej Sitarz \thanks{Partially supported by
KBN grant 2 P302 168 4} \thanks{E-mail: sitarz@if.uj.edu.pl} \\
Department of Field Theory \\
Institute of Physics \\
Jagiellonian University \\
Reymonta 4, 30-059 Krak\'ow, Poland}
\begin{titlepage}
\vspace{2cm}
\vfill
\maketitle
\begin{abstract}
We build a toy model of differential geometry on the real line,
which includes derivatives of the second order. Such construction
is possible only within the framework of noncommutative geometry.
We introduce the metric and briefly discuss two simple physical
models of scalar field theory and gauge theory in this geometry.
\end{abstract}
\vfill
\sc \noindent TPJU 2/94 \\
\sc January 1994
\vfill
\end{titlepage}

\section{Introduction}

The noncommutative differential geometry has proved
to be a very useful generalisation of the standard
differential geometry \cite{CON}-\cite{MAN}.
It enables us to construct and study models, which allows
description only in the language of algebras and modules.
However, often even in the classical well-known models we
are able to find new features or noncommutative
extensions \cite{JA1,JA2}.

This paper is devoted to the presentation of a toy model
of the differential algebra and sample physical models
on the real line (or circle), which, however, involve
the higher order derivatives. It is interesting that
such extension of the ordinary differential algebra,
demonstrates the noncommutative properties. We find it
similar to the situation we have discovered in the case
of discrete spaces, where the differential algebra was
noncommutative, thought the algebra of functions was
the commutative one.

In the paper we construct the differential algebra, then
we briefly discuss its metric properties and finally
we present a simple model of scalar field theory based
on the constructed geometry.

\section{The Differential Algebra}

Let $\a$ be the algebra of smooth functions on the real
line $C^{\infty}({\rb})$ with the pointwise multiplication and
addition. In the first step we shall construct a two-dimensional
bimodule over $\a$:
\begin{observation}
Let $dx$ and $\eta$ be the generators of the right module $\m$
over $\a$. Then, if we define the left multiplication rules
\bea
f dx & = & dx f + 2 \eta f', \l{de1} \\
f \eta & = & \eta f.  \l{de2}
\eea
where by $f'$ we denote the derivative of $f$, the module $\m$
shall be equipped with the structure of a bimodule.
\end{observation}
Now let introduce the operator $d$ acting on the algebra $\a$,
taking values in $\m$:
\be
df \= dx f' + \eta f''. \l{def1}
\ee
The operator $d$ has all properties of the external derivative,
as we shall demonstrate in the following lemma:
\begin{lemma}
The operator $d$ defined as above obeys the Leibniz rule:
\be
d (fg) \= (df) g + f (dg), \l{lem1}
\ee
and $\hbox{Ker}(d) = \cb$.
\end{lemma}
The last property follows directly from the definition (\ref{def1}),
therefore we shall concetrate only on proving (\ref{lem1}). By
definition we have:
\be
d (fg) \= dx (f'g + fg') + \eta (f'' g + 2 f' g' + f g''),
\l{lem1a}
\ee
where we have used the Leibniz rule for differentiating the
product of functions. Now, if we use the rules of the left
multiplication for the module $\m$ we shall obtain:
\bea
(df) g + f (dg) &=& dx (f'g) + \eta (f''g) + f dx g' + f \eta g'' \nonumber \\
& = &  dx (f'g + fg') + \eta ( f''g + 2 f'g' + f g''), \l{lem1b}
\eea
which is precisely the right-hand side of (\ref{lem1a}) and so
it ends the proof.

The bimodule $\m$ could be now seen as the module of one-forms,
and in order to construct the differential algebra we need to introduce
the product of one forms. First let us denote by $\Omega_0^n$ the
following direct sum:
\be
\Omega_0^n \= \oplus_{k=0}^n  \underbrace{
\m \otimes_\a \ldots \otimes_\a \m}_{\hbox{{\it n} times}},
\ee
where the first term (for $k=0$ is, of course, the algebra $\a$. We may
now denote the limit as $n \to \infty$ by $\Omega_0$. By definition,
$\Omega_0$ is an algebra, with the standard addition and the tensor
product defining the multiplication of its elements.

Now, let us define the ideal $\i \subset \Omega_0$ as the smallest
ideal generated by the elements $\eta \otimes \eta$ and
$\eta \otimes dx + dx \otimes \eta$. Now, we shall define our
differential algebra as the quotient $\Omega = \Omega_0/ \i$
and the product of forms, which we shall denote by $\bt$,
as the product in this quotient. In particular we have:
\bea
\eta \bt \eta & = & 0, \l{mu1} \\
\eta \bt dx & = & - dx \bt \eta. \l{mu2}
\eea

Our next task would bne the extension of the operator $d$
to the algebra $\Omega$ in such a way that it obeys the
graded Leibniz rule.
\begin{lemma}
Let us define the action of $d$ on $\m$ in the following way:
\be
d (f dx +  g \eta) = df \bt dx + dg \bt \eta + g dx \bt dx, \l{def3}
\ee
then $d$ has the following properties:
\bea
d^2 f = 0, \;\; \forall f \in \a, \l{dp1} \\
d (f \omega) = df \bt \omega + f (d\omega), \l{dp2} \\
d (\omega f) = (d \omega) f - \omega \bt df. \l{dp3}
\eea
\end{lemma}
The proof of the properties (\ref{dp1}-\ref{dp3}) follows directly
from the definitions (\ref{def1}),(\ref{def3}) and the rules of
left multiplications (\ref{de1},\ref{de2}).

Since we have already succesfully extended $d$ to $\m$ let us now
define $d$ on an arbitrary element of $\Omega_0$. We take:
\be
d (\omega_1 \ot \ldots \ot \omega_n) \=
\pi \left( \sum_{i=1}^n (-1)^{i+1}
\omega_1 \ot \ldots \ot (d\omega_i) \ot \ldots \omega_n \right){.}
\l{def5}
\ee
where $\pi$ is the qotient projection $\pi: \Omega_0 \to \Omega$.
In the next lemma we shall demonstrate that $d$ can be, in fact,
restricted to the algebra $\Omega$, thus the latter will be our
differential algebra.
\begin{lemma}
The ideal $\i$ is a differential ideal, i.e. the action of $d$ as
defined in (\ref{def5}) annihilates it:
\be
d \i \= 0,
\ee
so that $d$ may be defined on $\Omega$ and then it obeys
the graded Leibniz rule:
\be
d (\omega \bt \rho) \= (d \omega) \bt \rho +
(-1)^{\hbox{\footnotesize deg} \omega} \omega \bt (d \rho),
\ee for any $\omega,\rho \in \Omega$.
\end{lemma}
{}From the definition (\ref{def5}) it is clear that it is
sufficient to verify that $d \i_2 = 0$, where $\i_2$ is the
restriction of $\i$ to $\Omega^2_0$ and further we may
restrict only to its generators. From the definitions (\ref{def5})
and (\ref{def3}) we obtain:
\bea
d ( dx \ot \eta + \eta \ot dx) \= \pi \left( - 2 dx \bt dx \bt dx +
2 dx \bt dx \bt dx \right) \= 0, \\
d ( \eta \ot \eta ) \= \left( 2 dx \bt dx \bt \eta
- 2 \eta \bt dx \bt dx \right) \= 0,
\eea
which ends the proof of this part of the lemma. Now, since
$d$ is defined on $\Omega$, the graded Leibniz rule property
for the higher order forms follows directly from the definition
(\ref{def5}) and we already know that it is satisfied for $\a$
and $\Omega^1 = \m$, as demonstrated in (\ref{lem1}) and
(\ref{dp2},\ref{dp3}). Of course, it is nilpotent, i.e. $d^2=0$,
which is the consequence of (\ref{dp1}) and the graded Leibniz
rule.

The above constructed differential algebra is infinite-dimensional and
each its component (the module of $n$-forms for any fixed $n>0$)
is a two-dimensional free bimodule over $\a$. The standard,
finite differential algebra on ${\rb}$ is, of course,
its subalgebra.

The diffeomorphism $x \to y(x)$ of ${\rb}$ generates the following
automorphism of $\m$:
\bea
\eta & \to & \eta \frac{dy}{dx}, \\
dx & \to & dx \frac{dy}{dx} + \eta \frac{d^2y}{dx^2}.
\eea

\section{Metric}

Before we proceed with the introduction of the analogue of
metric we shall prove the following simple lemma:
\begin{lemma}
There exists a conjugation $\star$ on the algebra $\Omega$, which
after restriction to $\a$ is just the complex conjugation and
it graded commutes with $d$:
\be
d (\omega^\star) \= (-1)^{\hbox{\footnotesize deg} \omega} (d \omega) ^\star.
\ee
\end{lemma}
The proof is straightforward if we take $f^\star = \bar{f}$ for
$f \in \a$ and $dx^\star = dx$ and $\eta^\star = - \eta$.

We define the metric as a middle linear functional \cite{JA1}-\cite{JA3}:
\bea
G: \m \otimes_\a \m \; \to \; \a, \l{c1} \\
G(a \omega b, \rho c) \= a G(\omega, b \rho) c. \l{c2}
\eea
The condition (\ref{c2}) is very strong and in particular, it
results in the following constraints:
\bea
G(\eta,\eta) & = & 0, \\
G(dx, \eta) & = & - G(\eta, dx).
\eea
We may also require that the metric is hermitian, i.e.:
\be
G(\omega, \rho) \; = \; \left( G(\rho^\star,\omega^\star) \right)^\star,
\ee
and then we obtain that both $G(dx,dx)$ and $G(\eta,dx)$ must
be real.

As the trace on the algebra $\a$, we take the diffeomorphism
invariant integral:
\be
\hbox{Tr} f \= \int_{{\rb}} dx \frac{1}{\sqrt{G(dx,\eta)}} f(x),
\ee
with the diffeomorphism invariant measure $dx \frac{1}{\sqrt{G(dx,\eta)}}$.

\section{Field Theory}

Let us construct now two simple models of field theory in the
considered geometrical setup. We shall begin here with the
siplest one, the scalar field theory.

\subsection{Scalar Field Theory}

Let $\phi$ be the scalar field, which is just the element of $\a$.
We define the action ${\s}(\phi)$ in the simplest way:
\be
\s(\phi) \= \hbox{Tr} G(d\phi^\star, d\phi). \l{sf1}
\ee
Now, if we write the right-hand side of (\ref{sf1}) explicitely
we obtain:
\be
\s(\phi) \= \int_{\rb} dx \frac{1}{\sqrt{G(dx,\eta)}} \left( G(x,x)
\phi' \bar{\phi}' + G(dx,\eta) (\bar{\phi}'' \phi' +
\bar{\phi}' \phi'') \right){.} \l{sf3}
\ee
Now, if denote by $F(x)$ such function that
$F' = \sqrt{G(dx,\eta)}$ we may rewrite (\ref{sf3}) as:
\be
\s(\phi) \= \int_{\rb} dx \frac{1}{\sqrt{G(dx,\eta)}} \left(
G(x,x) - F(x) \sqrt{G(dx,\eta)} \right) \left( \bar{\phi}' \phi'
\right){.} \l{sf4}
\ee
where we have eliminated the boundary terms (full derivative
of $\sqrt{G(dx,\eta} \bar{\phi}' \phi'$).

Now, this does not differ from the clasical action of the scalar
field in one-dimension (up to the boundary terms, of course),
with the metric $E$:
\be
E(dx,dx) \= \frac{G(x,x) - F(x)\sqrt{G(dx,\eta)} }{ \sqrt{G(dx,\eta)} },
\ee.
so the effective theory would be the same, the only possible difference
arising from the above mentioned boundary terms.

\subsection{Gauge Theory}

Here, we would like to present the possibility of construction
of a simple $U(1)$ gauge theory in the considered differential
geometry. This is a completely new feature, as in the standard
differential geometry one cannot build gauge theory in less
than two dimensions.

The gauge group ${\cal G}$ shall be in our case the unitary group of $\a$
consisting of all $U(1)$ valued functions on $\rb$, and the hermitian
connection will be the one-form $A$:
\be
A \= dx A_x + \eta A_\eta,
\ee
such that $A^\star = - A$, which is satisfied only if
\bea
\bar{A_x} & = & - \bar{A_x}, \l{er1}\\
\bar{A}_\eta  & = & A_\eta - 2 A_x'. \l{er2}
\eea
Their gauge transformations rules are as follow:
\bea
A_x \to  A_x + i \phi', \\
A_\eta  \to  A_\eta  - 2 i A_x \phi' + i \phi'' + (\phi')^2,
\eea
where $\phi$ is a real field defining $e^{i\phi} \in U(\a)$.
{}From the conditions (\ref{er1},\ref{er2}) we see that in fact
only the imaginary part of $A_x$ and real part of $A_\eta$ make
independent real variables. If we call them $\Phi$ and $\Psi$,
respectively, we may rewrite the connection as:
\be
A \= dx (i \Phi) + \eta (\Psi + i \Phi'),
\ee
with the tranformation rules:
\bea
\Phi \to \Phi + \phi', \\
\Psi \to \Psi + 2 \Phi \phi' + (\phi')^2.
\eea
Let us notice that by gauge transformation we can eliminate $\Phi$
or in some cases $\Psi$ but not both of them at the same time.

Finally, let us calculate the curvature $F=dA + A \bt A$:
\bea
F & = & dx \bt dx \left( \Psi - \Phi^2 \right) \\
 & + & dx \bt \eta \left( \Psi' - 2 \Phi \Phi' \right){.}
\eea
The curvature is, of course, selfadjoint, i.e. $F^\star = F$
and additionally we have $dF=0$.

There exist two possible gauge invariant lagrangian
terms, one linear in $F$:
\be
{\cal L}  \sim  \left( \Psi - \Phi^2 \right),
\ee
since by the similar argument as in the case of a scalar field
we can eliminate the second component of $F$, which effectively
gives contribution to some boundary terms. The other one
shall be simply its square:
\be
{\cal L}  \sim  \left( \Psi - \Phi^2 \right)^2, \l{pot}
\ee
and it appears that it is equal to the Yang-Mills lagrangian
in our case.

Both of the derived candidates for the Lagrange function
contain no dynamical dependance on the fields $\Phi$
and $\Psi$, therefore the resulting gauge theory is trivial.
Let us notice, however, that in combination with the
standard gauge theory on a smooth manifold it could provide
the potential term for two real fields, which then could
be treated also as fields on the manifold. The potential
(\ref{pot}) is rather similar to the Higgs potential, so
the above construction may provide a similar symmetry-breaking
mechanism as one obtains in the case of discrete geometry.

\section{Conclusions}

As we have demonstrated in this paper one could extend the
differential calculus on the real line to include derivatives
of second order, the construction appears, however, to
be a noncommutative one. Of course, following the same procedure
as described in the second section we may generalise the
idea to higher order derivatives and by taking tensor products
of the obtained differential algebras we may also build such
calculus for $\rb^n$, however, further generalisation to arbitrary
smooth manifolds would be more difficult.

What is a remarkable picture of this geometry, is that
due to its noncommutative properties we obtain no new physics
if we attempt to construct physical models. Naively, one would
expect that such models would include higher order terms
like $\phi''\bar{\phi}''$. In our case, the only
difference with the classical geometry appears in the
existence of possible topological (or boundary) terms. The
gauge theory, though possible to construct, has no dynamical
degrees of freedom and may only contribute to some potential
terms in the products with some standard gauge theory.

We may believe that this could be a profound geometrical
reason that no higher order derivatives are necessary in
the description of fundamental physical theories.

Although we treat the above construction mainly as a toy model
there are still some interesting problems in it. First, one
could study such differential algebras on smooth manifolds
in general. From the physical point of view, it would be
interesting to analise the gauge theories on the product of
standard geometry and the one discussed above.

\end{document}